\documentclass[oldversion]{aa}
\pdfoutput=1

\usepackage{ifthen}
\usepackage{graphicx}
\usepackage{natbib}
\usepackage{textcomp}
\usepackage{amsmath}
\usepackage{amssymb}
\usepackage{wasysym}
\usepackage{fancybox}
\usepackage{multirow}

\newboolean{isReferee}
\setboolean{isReferee}{false}

\newcommand{\td}[0]{\textdegree}
\newcommand{\co}[0]{CoRoT}

\newcommand{\fe}[0]{pdf}

\begin{document}

\title{A planetary eclipse map of CoRoT-2a}
\subtitle{Comprehensive lightcurve modeling combining rotational-modulation and transits}

\author{K. F. Huber
   \and S. Czesla
   \and U. Wolter
   \and J. H. M. M. Schmitt}

\institute{Hamburger Sternwarte, Universit\"at Hamburg, Gojenbergsweg 112, 21029 Hamburg, Germany}

\date{Received ... / Accepted ... }

\abstract {We analyze the surface structure of the planet host star CoRoT-2a using a consistent model for
           both the `global' (i.e., rotationally modulated) lightcurve and the transit lightcurves, 
           using data provided by the CoRoT mission.
           Selecting a time interval covering two stellar rotations
           and six transits of the planetary companion \mbox{CoRoT-2b},
           we adopt a `strip' model of the surface to reproduce the photometric modulation inside and outside the
           transits simultaneously.
           Our reconstructions show that it is possible to achieve appropriate fits for the entire sub-interval
           using a low-resolution surface model with $36$~strips.
           The surface reconstructions
           indicate that the brightness on the eclipsed section of the stellar surface is \mbox{($6\pm1$)~\%} lower than
           the average brightness of the remaining surface.
           This result suggests a concentration of stellar activity
           in a band around the stellar equator similar to the behavior observed on the Sun.}


\maketitle

\section{Introduction}
\label{Sec:Intro}

Astronomers have long been interested in the surface structure of active stars and their evolution.
Yet, the surfaces of stars other than the Sun can hardly be resolved directly, so that indirect techniques must be
used to obtain an image of the surface.
One such technique is Doppler imaging \citep{Vogt1983}, which requires a dense series of high resolution spectra
and stellar rotation velocities of \mbox{$v\sin(i)\,\apprge\,20$~km/s}
(compared to $v_{eq}\,\approx\,2$~km/s for the Sun).
Alternatively, lightcurves also yield information on stellar surface structures and can usually be obtained
at low observational cost.
However, photometry provides less information and the problem of lightcurve inversion is known to be notoriously ill-posed.

Since the launch of CoRoT in 2006, an increasing amount of high quality space-based photometry has become available.
Without the limitations the atmosphere and the day-night cycle impose on ground based observatories,
CoRoT is able to provide photometry with unprecedented temporal coverage and cadence,
which is enormously interesting in the context of stellar activity and surface reconstruction.

In the course of the CoRoT planet hunting project, the giant planet CoRoT-2b \citep{Alonso2008} was detected.
The host star of this planet, CoRoT-2a, is solar like in mass and radius, but rotates approximately four times
faster than the Sun and is considerably more active.
The planet orbits its host star approximately three times per stellar rotation and,
during its passage across the stellar disk, acts as a shutter scanning the surface of the star along a well defined latitudinal band.
As the `local' surface structure is imprinted on the transit profiles \citep{Wolter2009, Czesla2009},
they can be used to partially resolve the ambiguity of the lightcurve inversion problem.

While \citet{Lanza2009} used the `global' lightcurve of the host star
to reconstruct its surface inhomogeneities, without
considering the transits,
\citet{Wolter2009} concentrated on a single transit lightcurve to reconstruct a fraction of the surface, neglecting the `global' lightcurve.
In this work, we combine and refine these approaches to present a reconstruction which simultaneously describes both the overall
lightcurve and the transits during two stellar rotations.

\section{Observations and data reduction}
\label{Sec:Object}

\citet{Alonso2008} discovered the planet CoRoT-2b using photometric data provided by the CoRoT mission
\citep[for a detailed description, see][]{Auvergne2009}.
The planet was detected in the field observed during the first \textit{long run} carried out between May~$16^\mathrm{th}$ and \mbox{Oct. $15^\mathrm{th}$ 2007}.
The default sampling rate of \co~photometry is $1/512$~s$^{-1}$.
The \co-2 lightcurve was observed at this rate only for the first five days, after
which the transits were detected and the satellite switched to alarm-mode, now taking data every $32$~s.
The light collected by the \co~telescope is dispersed using a prism and recorded by a CCD chip.
Individual sources are separated by a photometric mask, which also defines three broad-band channels (nominally red, green, and blue).
Currently, there is no appropriate calibration available for these channels, so that it is unfeasible to use the color
information in this work.
The signal obtained by summing up the individual channels, often referred to as `white light',
corresponds to an optical measurement with a filter transmission maximum in the red wavelength region \citep{Auvergne2009}.
Accordingly, \citet{Lanza2009} assume an isophotal wavelength of $700$~nm for their pass-band.
The \co~data undergo a standard pipeline processing, during which
data points that are significantly affected by known events,
as for example the South Atlantic Anomaly (SAA), are flagged, so that they can be removed
from the lightcurve.

The host star \co-2 has a spectral type of G7V with an optical companion at a distance of approximately \mbox{$4.3$''}
\citep[2MASS,][]{Skrutskie2006}, too close to be resolved by CoRoT.
According to \citet{Alonso2008} the secondary contributes
a constant fraction of $(5.6 \, \pm 0.3)$~\% to the total CoRoT-measured flux.
In Table~\ref{Tab:Exo2prop} we list the system parameters of CoRoT-2a/b, which are used throughout our analysis.
CoRoT-2b's orbital period of \mbox{$\approx 1.74$~days} is about a third of CoRoT-2a's rotation period.
Hence, the  almost continuous CoRoT data sample of $142$~days covers about $30$~stellar rotations and more than  $80$~transits.
The lightcurve shows signatures of strong stellar activity and substantial rotational modulation \citep{Lanza2009}.
We use the same CoRoT raw data reduction procedures as described in \citet[Sect.~2]{Czesla2009}.


\begin{table}[t!]
  \begin{minipage}[h]{0.5\textwidth}
    \renewcommand{\footnoterule}{}
    \caption{Stellar/planetary parameters of CoRoT-2a/b. \label{Tab:Exo2prop}}
    \begin{center}
      \begin{tabular}{l c c}
      \hline \hline
      Star \hspace*{0.0cm} \footnote{$P_s$ - stellar rotation period, $P_s^*$ - stellar rotation period used for the observation interval analyzed in this paper (see Sect.~\ref{Sec:RotPer}).}  & Value $\pm$ Error & Ref.\footnote{taken from \citet{Lanza2009} [L09], \citet{Alonso2008} [A08], \citet{Bouchy2008} [B08], or \citet{Czesla2009} [C09]} \\
      \hline
      $P_s$          & $(4.522 \, \pm \, 0.024)$ d         & L09 \\
      $P_s^*$        & $4.57$ d                            &  \\
      Spectral type  & G7V                                 & B08 \\
      \hline \hline
      Planet \hspace*{0.0cm} \footnote{$P_p$ - orbital period, $T_c$ - central time of first transit, $i$ - orbital inclination, $R_p, R_s$ - planetary and stellar radii, $a$ - semi major axis of planetary orbit, $u_a, u_b$ - linear and quadratic limb darkening coefficients.}  & Value $\pm$ Error & Ref.\\
      \hline
      $P_p$       & $(1.7429964 \, \pm \, 0.0000017)$ d   & A08 \\
      $T_c$ [BJD] & $(2\,454\,237.53362 \, \pm \, 0.00014)$ d & A08 \\
      $i$            & $(87.7 \, \pm \, 0.2)$\textdegree & C09 \\
      $R_p/R_s$   & $(0.172 \, \pm \, 0.001)$           & C09 \\
      $a/R_s$     & $(6.70 \, \pm \, 0.03)$               & A08 \\
      $u_a, u_b$      & $(0.41\pm0.03), (0.06\pm 0.03)$           & A08 \\
      \hline
      \end{tabular}
    \end{center}
  \end{minipage}
\end{table}

\section{Analysis}
\label{Sec:Analysis}

\subsection{Modeling approach}
The measurements of the Rossiter-McLaughlin effect by \citet{Bouchy2008} suggest that the rotation axis of the host star and
the planet's orbit normal are approximately co-aligned \mbox{($\lambda \, = \, 7.4 \, \pm \, 4.5$\textdegree)}.
$\lambda$ represents the misalignment angle projected on the plane of the sky, and its value
strongly favors aligned orbital and rotational axes, even though it does not prove it.
Further support for a co-aligned geometry comes from the following argument:
Comparing the measured \mbox{$v\sin(i)\,=\,11.85$~km/s} with a calculated equatorial velocity of
\mbox{$v_{eq}\,=\, 2\pi R_s / P_s \,\approx\, 10$~km/s} derived with the theoretically obtained value
\mbox{$R_s\,=\,0.9\cdot R_\odot$} \citep{Alonso2008} also favors
\mbox{$\sin(i)\,\approx\,1$}.

Therefore, the planet always eclipses the same low-latitude band between $6$ and $26$~degrees.
The transits separate the stellar surface into two observationally
distinct regions, i.e., a region eclipsed by CoRoT-2b and another region that is not.
In the case of CoRoT-2a the eclipsed section covers $\approx 21$~\% of the stellar disk
corresponding to $\approx 17.3$~\% of its surface.
The time-resolved planet migration across the visible stellar disk sequentially covers and uncovers surface fractions,
so that the brightness profile of the underlying stellar surface is imprinted on the transit lightcurve.
  
  \begin{figure}[b]
    \centering
    \includegraphics[width=0.35\textwidth]{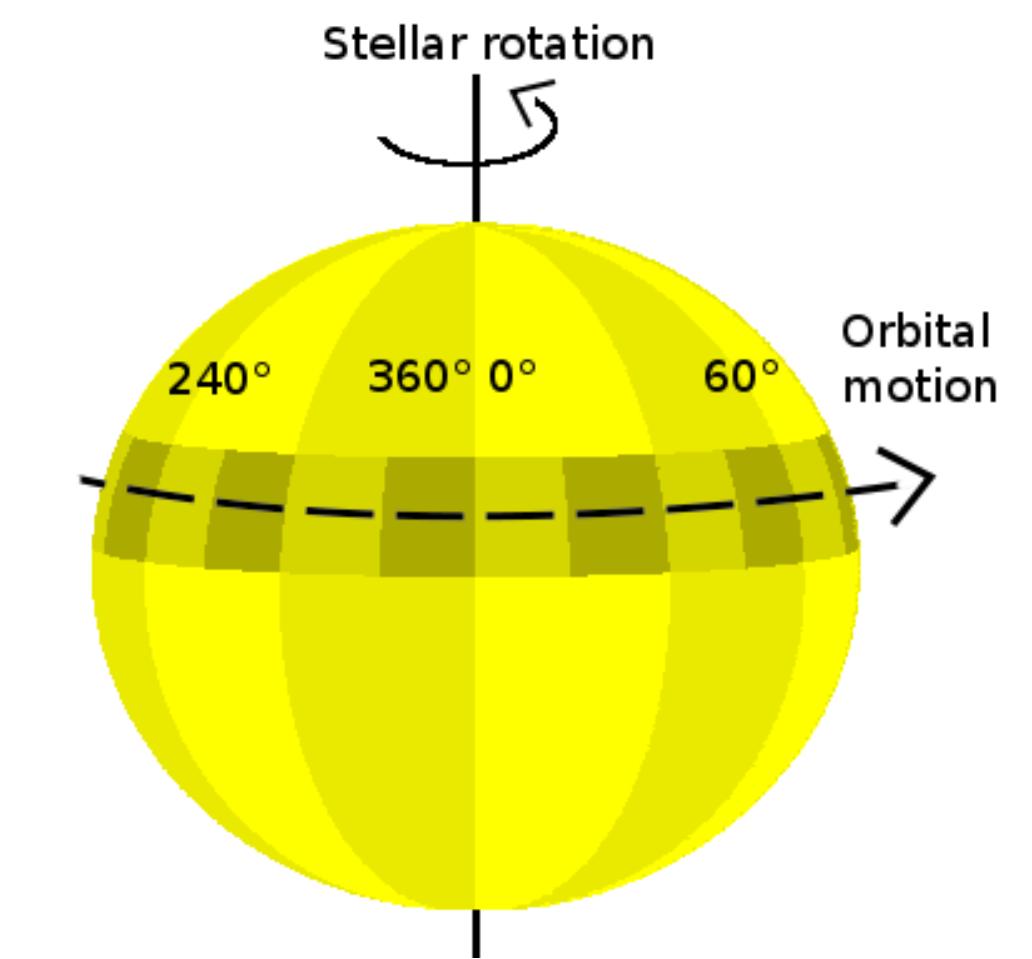}
    \caption{Our model geometry using $12$~longitudinal strips for the non-eclipsed and $24$~strips for the 
    eclipsed section, respectively.
    \label{fig:ModGeom}}
  \end{figure}
  
For our modeling, we separate the surface into the eclipsed and the non-eclipsed section, which are both further subdivided
into equally sized, longitudinal bins or `strips' as demonstrated in Fig.~\ref{fig:ModGeom}. Let $N_{e}$ be the number of
bins in the eclipsed section and $N_{n}$ be the number of non-eclipsed strips.
As is apparent from Fig.~\ref{fig:ModGeom}, $N_{e}$ and $N_{n}$ need not be the same.
Altogether, we have $N_{tot} = N_{e} + N_{n}$ bins enumerated by some index $j$.
To each of these surface bins a brightness $b_{j}$ is assigned, with which it contributes to the total (surface) flux of the star.
Now let $V_{ji}$ denote the visibility of the $j$-th bin at time $t_i$.
The visibility is modified in response to both a change in the viewing geometry caused by the stellar rotation and a transit of the planet.
The modeled flux $f_{mod,i}$ at time  $t_{i}$ is then given by the expression
\begin{equation}
  f_{mod,i} = \sum_{j=1}^{N_{tot}} V_{ji} b_{j}\, .
  \label{eq:ModelFlux}
\end{equation}
We determine the unknown brightnesses, $b_{j}$, by comparing $f_{mod,i}$ to a set of $M_C$ 
CoRoT flux measurements using a specifically weighted version of the $\chi^2$-statistics:
\begin{equation}
  \chi^2_{m} = \sum_{i=1}^{M_C} \frac{(f_{mod,i} -f_{obs,i})^2}{\sigma_i^2}\cdot w_i \ .
\end{equation}
$\chi^2_m$ differs from $\chi^2$ by a weighting factor, $w_i$, which we choose to be $10$ for lightcurve
points in transits and $1$ otherwise; in this way
the global lightcurve and the transits are given about the same priority in the minimization process.
Error bars for the individual photometric measurements were estimated from the data point distribution
in the lightcurve, and the same value of \mbox{$\sigma=1000$~e$^-$/32~s} ($=1.4\times 10^{-3}$ after lightcurve normalization)
was used for all points.

In our modeling we currently exclude surface structures with a limb-angle dependent contrast; this particularly
refers to solar-like faculae, for which \citet{Lanza2009} find no evidence in their analysis.
The planet, CoRoT-2b, is modeled as a dark sphere without any thermal or reflected emission.
This approximation is justified by the findings of \citet{Alonso2009}, who report a
detection of the secondary transit
with a depth of \mbox{$(0.006 \, \pm \, 0.002)$~\%},
which is negligible in our analysis.

The actual fit is carried out using a (non-gradient) Nelder-Mead simplex algorithm (e.g. \citeauthor{NR-BOOK}~1992).
In our model all strips are mutually independent, and as we define only a rather coarse strip subdivision for the
non-eclipsed section of the surface and
since the eclipsed section is thoroughly covered by the transits, no further regularization is necessary.

\subsection{Fit parameter space - parametrization, restrictions, and interpretation}

Our fit space has a total of $N_{tot} = N_{e} + N_{n}$ dimensions,
and the associated parameters are the brightnesses, $b_{j=1..N_{tot}}$.
The most obvious choice of fit parameters are the brightnesses themselves.
Yet, using a slightly different definition in our algorithm provides some advantages.
Instead of using the brightness of the global strips in our fits, we
replace them with a weighted sum of the brightnesses of the eclipsed and the non-eclipsed strips covering the same
longitudes.
This quantity $z$ is a measure of the total flux emitted from all strips enclosed within a certain
longitude range and, therefore, represents the level of the global lightcurve independent of how the brightness is
distributed among the individual strips contributing to the sum.
Without any transit observations the distribution of flux
among the individual contributors could hardly be further restricted, because latitudinal information could not
be recovered.
Thus, we use the tuple
$(b_{l=1..N_{e}}, z_{k=1..N_{n}})$ for our reconstructions,
where $z$ is defined by
\begin{equation}
  z_k = b_{N_e+k} + \frac{1}{c \cdot q}\sum_{s=q_0}^{s<q_0+q} b_s \, .
  \label{eq:z}
\end{equation}
In Eq.~\ref{eq:z}
$b_{N_e+k}$ denotes the brightness of the k-th global strip, $q$ is defined by
$N_e/N_n$ (the factor by which the eclipsed section is oversampled compared to the non-eclipsed section),
the index range $q_0 \le s < q_0+q$ enumerates all eclipsed strips covering the same longitudes as the global strip referred to by $b_{N_e+k}$,
and $c$ is a scaling factor accounting for the size difference between the eclipsed and the non-eclipsed section.

The practical advantage of using $z$ instead of the brightness values themselves lies in the parameter interdependence.
Assume a fit algorithm adjusts the structure of
a transit lightcurve using the eclipsed strips; every modification of their brightness causes a modification of
the global lightcurve level which must possibly be compensated by an appropriate adjustment of the global strip's brightness. Such an
adjustment is inherent in the definition of $z$, so that $b_{l=1..N_e}$ and $z_{k=1..N_n}$ become largely independent quantities.
In our fits we use $c=5$, which roughly corresponds to the ratio of disk area covered by global and eclipsed strips.

To normalize the observed \co-2a lightcurve, we divided all measurements by the largest flux value in our lightcurve
so that \mbox{$0<$~(normalized flux)~$\le 1$}. The matrix $V_{ji}$ in Eq.~\ref{eq:ModelFlux} is normalized
according to
$$
  \sum_{j=1}^{N_{tot}} V_{ji} = 1 \; \mbox{for all $i$} \; ,
$$
which yields $f_{mod,i} = 1$ for $b_j=1$, i.e., a constant model lightcurve at level $1$. In a first, tentative interpretation a
surface element with the brightness~$1$ corresponds
to a photospheric element free of any spots. Yet, this is only correct as long as we assume that the largest observed flux in the lightcurve,
indeed, represents the `spot-cleaned' photospheric luminosity. As \co-2a is, however, a very active star, it seems
probable that polar spots persist on its surface. Moreover, it seems likely that lower latitude structures cover
a fraction of the stellar surface even if the lightcurve is at maximum.
For this reason, individual surface elements (strips) may be brighter than the `average surface' during the maximum observed flux.
While such information could not be recovered if no transits were observed, individual surface regions eclipsed by the planet
can conceivably be brighter than the `global' photosphere seen during lightcurve maximum.
Therefore, we do not exclude strips with brightness values larger than $1$ in our fits,
i.e., we do not fix the photospheric brightness;
this results in brightnesses greater than $1$ for individual strips (e.g. Fig.~\ref{fig:modelSurface}).
The only parameter space restriction applied during our fits is that the brightness must be positive.

\subsection{Which part of the lightcurve should be used?}

In order to derive a meaningful model we need to select a time span,
which is both long enough to provide an appropriate coverage of the surface, and short enough
to minimize the effects of surface evolution;
the latter, while doubtlessly present, appears slow compared to the stellar rotation period.
\citet{Lanza2009} give typical lifetimes of $55$~d ($\approx 12$~rotations) for active regions and $20-30$~d for some
individual spots.
In our analysis, we use the time span ranging from phase $1.85$ through $3.85$
(\mbox{$\mathrm{BJD} = 2\,454\,245.988$} to \mbox{$\mathrm{BJD} = 2\,454\,255.128$},
\mbox{$\mathrm{BJD}$\,=\,Barycentric Julian Date}),
which covers $6$~transits and shows only little variation in the global lightcurve.
The data are re-binned using a binsize of $128$~s for the transit covered periods and $2\,016$~s for the remaining lightcurve.

Our binning approach has to take into account interruptions of the lightcurve due to data drop outs (for instance due to the
South Atlantic Anomaly) and, of course, has to account for the change in bin size when a transit period begins or ends. Moreover,
the \co-2 lightcurve is sampled at two different rates ($1/512$~s$^{-1}$ and $1/32$~s$^{-1}$), which does, however, not impose
a problem during the time span under consideration here. To obtain
the binned curve, we average all flux values comprised by a bin and place the resulting value at the barycenter of the
contributing time stamps. To compute the error, we divide the standard deviation for individual points by the square root of
the number of averaged points. With this approach we (typically) obtain an error of $7\times 10^{-4}$ for in-transit points and
$1.8\times 10^{-4}$ for out-of-transit points.

\begin{figure}[b]
    \includegraphics[angle=-90, width=0.49\textwidth]{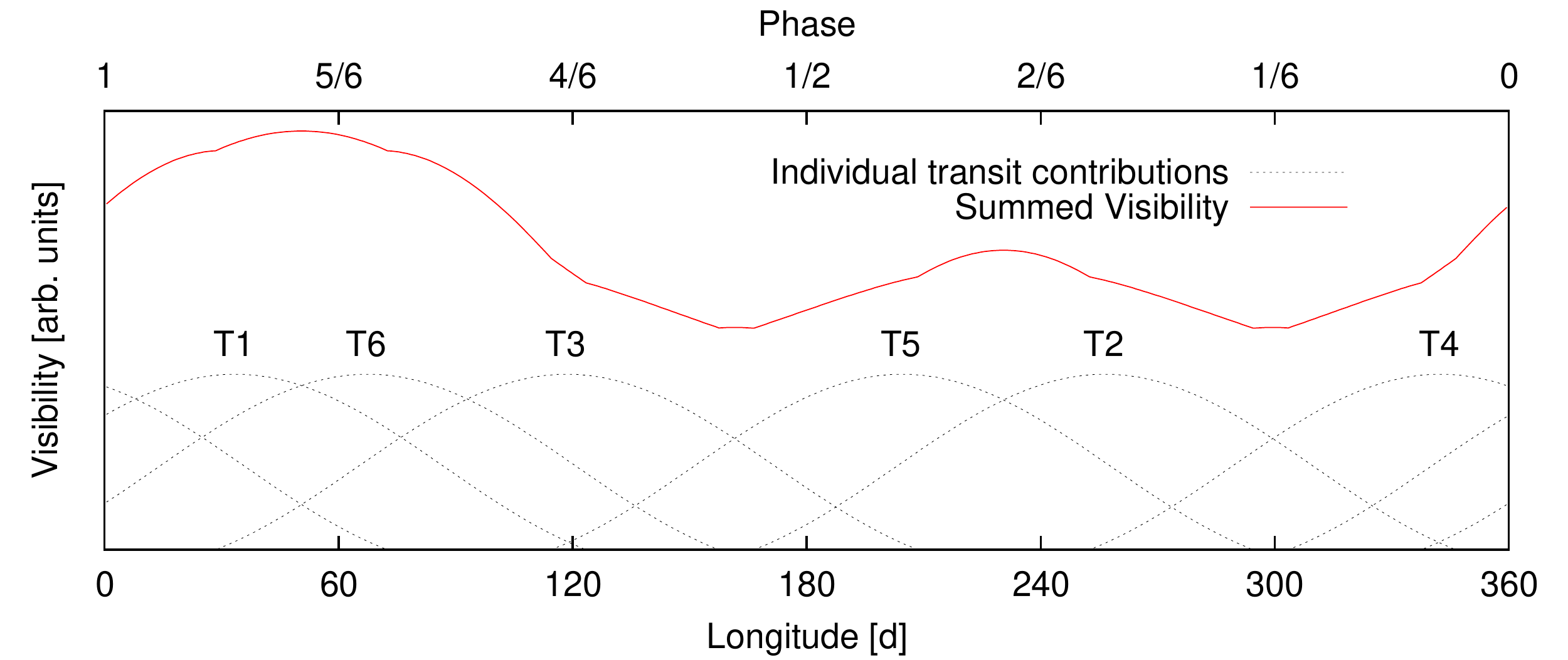}
    \caption{Visibility of the eclipsed stellar surface during the transits in the selected observation interval.
             A low visibility means that a stellar feature at the corresponding longitude has a low impact on
             the transit profiles.
    \label{fig:PhVis}}
  \end{figure}

  In Fig.~\ref{fig:PhVis} we demonstrate the coverage of the eclipsed surface section by these $6$~transits within the selected phase interval.
  A single rotation phase including three transits provides only a very inhomogeneous
  `scan' of the eclipsed surface due to limb darkening, projection geometry, and the distribution of transit intervals
  (cf. Fig.~\ref{fig:PhVis}).
  As a transit occurs every $\approx 0.4$~stellar rotations, a homogeneous coverage of one full rotation
  is achieved using five transits.
  Nonetheless, we decide to use an integer number of stellar rotations and use six transits
  with the last one showing virtually the same part of the eclipsed surface as the first.
  
\subsection{Surface evolution, rotation period, and model limits}
\label{Sec:RotPer}
Although the lightcurve of \co-2a shows remarkably periodic minima and maxima, the rotation period of
the star is not exactly known.
Using a Lomb-Scargle periodogram, \citet{Lanza2009} find a
rotation period of $(4.52\pm0.14)$~d for the star, which is further refined in the course of their surface modeling.
Assuming that the longitudinal migration of the active longitudes should be minimal, \citeauthor{Lanza2009}
pin down the stellar rotation period to $4.5221$~d.
While this rotation period minimizes the migration of the active longitudes,
it results in individual spots showing a retrograde migration with an apparent
angular velocity $\approx 1.3$~\% lower than the stellar rotation.

For our modeling we determine an `effective' period,
representing the rotation period of the dominating surface features we are mainly interested in.
In our approach, we use the selected part of the lightcurve, remove the transits,
and fold the remaining lightcurve back at a number of periods between $4.4$~d and $4.7$~d.
The best match is obtained using a period of $4.57$~d, which also results in the best fits of our models;
therefore, we will use it throughout our analysis.
This period is also in agreement with the values given by \citet{Lanza2009} assuming a rotation period of $4.5221$~d and
a mean retrograde migration `slowing down' the spots by $1.06$~\%.
Note, however, that changes of the rotation period on this scale do not result in significantly different surface reconstructions.

Even though we identified a lightcurve interval
with relatively weak surface evolution,
and refine the rotation period to account for some evolutionary effects,
there is still a remaining modulation.
The presence of this modulation imposes a principal limit on the fit quality
that can be achieved by adopting a static model to the lightcurve, because both
stellar rotations have to be described by the same model.
To estimate this limit, again for the global lightcurve only, we estimated the quantity
\begin{equation}
  <\Delta \chi^2> = \frac{1}{N} \sum_{i=1}^{N} \left( \frac{f(p_i)-f(p_i+1.0)}{2\sigma} \right)^2 \approx 14.2 \; .
  \label{eq:DeltaChiSqr}
\end{equation}
Here, $f(p_i)$ is the normalized flux in the i-th phase bin, $f(p_i+1.0)$ is the flux measured at the same phase during
the next stellar rotation, and the sum stretches over all phases pertaining to the first rotation.
$f(p_i+1.0)$ was obtained by interpolation, because the phase sampling is not exactly the same in both rotations.
Since $\sigma_i \approx \sigma$, the best conceivable common model with respect to $\chi^2$
at phase point $p_i$ is given by $(f(p_i)+f(p_i+1.0))/2$, and, therefore, the sum in Eq.~\ref{eq:DeltaChiSqr} estimates
the $\chi^2$ contributions induced by surface evolution for each point.
If there was no surface evolution, the expression in Eq.~\ref{eq:DeltaChiSqr} would equate to $0.5$,
because statistical errors are, of course, still present.
Therefore, a limit of $\chi^2 \approx 14.2$ per (global) lightcurve point will not be overcome by any static model.
Equivalently, the expectation value, $<\Delta f/2>$, for the flux deviation from the best model equates to
$<\Delta f/2> = <(f(p_i)-f(p_i+1.0))/2> = 5.6\times 10^{-4}$ and cannot be surpassed.

\subsection{Model resolution}

  The parameters $N_{e}$ and $N_{n}$ specify the model resolution of the eclipsed and non-eclipsed sections.
  An appropriate choice of these parameters balances fit quality and model ambiguity;
  this way the largest possible amount of information can be extracted.
  
  In order to find the optimal value for the number of non-eclipsed strips, we carry out fits
  to only the global lightcurve using an increasing number of global strips.
  Starting with only $4$~strips, we find the reduced $\chi^2$ value, $\chi^2_R$, to decrease rapidly until $8$~strips are used.
  From this point, $\chi^2_R$ responds only weakly to an increase of the strip number, but still decreases. Using $12$
  strips, we find $\chi^2_R = 16$.
  With an estimated `socket' contribution of $\approx 14.4$ provided by surface evolution, we
  attribute a fraction of $\chi^2_R\approx 1.6$ to statistical noise. This fraction decreases to $\approx 1$ if we use
  $30$~strips, in which case we obtain a longitudinal resolution of $12$\textdegree, comparable to that achieved by \citet{Lanza2009}.
  According to our test runs, we obtain reasonably stable results using $12$~strips.
  As the stability of the solutions decreases for larger strip numbers, while $\chi^2_R$ only slightly improves,
  we argue in favor of using $12$ global strips in our modeling, to extract the largest possible amount of physically relevant results.

  The resolution used on the eclipsed surface band, is determined according to the following considerations.
  The extent of the planetary disk on the center of the stellar disk is about $20$\textdegree$\times20$\textdegree. All stellar
  surface elements simultaneously (un)covered by the planet's disk are equivalent with respect to
  our lightcurve modeling. Individual features can, thus, be located (or smeared out) all along the edge of the planetary
  disk to provide the same effect in the lightcurve. This edge stretches across $10$\textdegree \ in longitude (only the `forward'
  part) and $20$\textdegree \ in latitude, which defines a fundamental limit for the resolution.
  Assuming a particular shape for the features, decreases the degree of ambiguity as was for example shown by
  \citet{Wolter2009}.
  
  A meaningful structure in the transit profile should comprise at least $3$ consecutive
  lightcurve bins corresponding
  to about $360$~s or $\approx 6$\textdegree \ of planet movement across the center of the stellar disk.
  The extent of individual strips should, therefore, not fall below this limit;
  it should even be larger.
  
  Combining these arguments with the results of our test runs, we decided to use
  $24$~strips on the eclipsed section, so that a longitudinal resolution of $15$\textdegree \ is achieved. With this choice,
  a single strip on the eclipsed band appears about the same size (face-on) as the planetary disk. Additionally, we note that
  this approximately corresponds to the resolution used by \citet{Lanza2009} in their maximum entropy reconstructions.

\subsection{Results of the modeling}

  In our analysis, we achieve a longitudinal resolution of $\approx 15$\td \ on the eclipsed section
  making up $\approx 17$~\% of the stellar surface
  and $30$\td \ for the rest.

\begin{figure}[b]
    \includegraphics[angle=-90, width=0.49\textwidth]{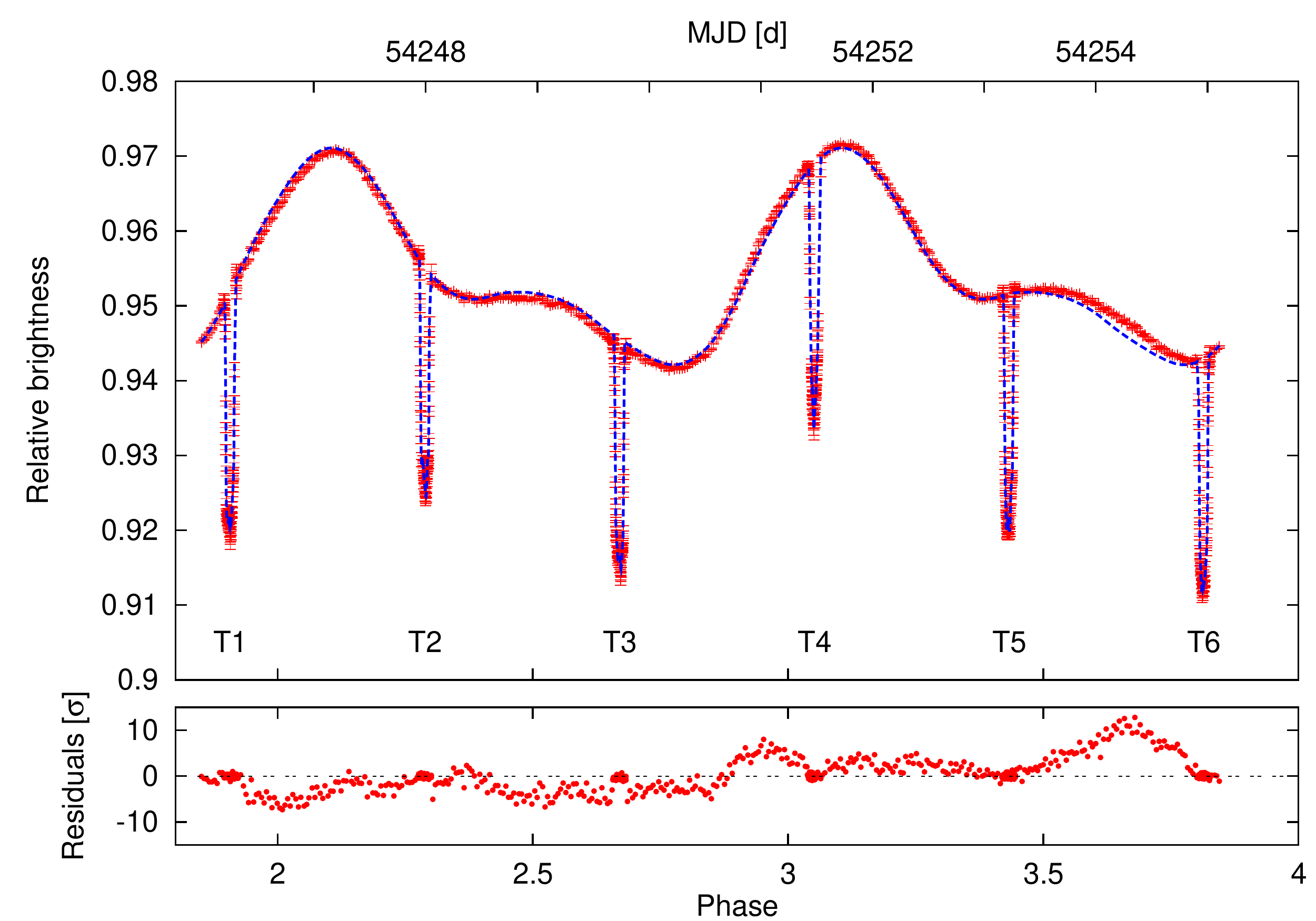}
    \caption{Upper panel: CoRoT data for rotational phases $1.85$ to $3.85$
             ($2\,016$~s binning for the global lightcurve and $128$~s for the transits, red symbols)
             and our model lightcurve (dashed blue curve).
             Lower panel: Residuals of our model.
    \label{fig:modelLC_global}}
  \end{figure}

\begin{figure}[t]
    \includegraphics[angle=-90, width=0.49\textwidth]{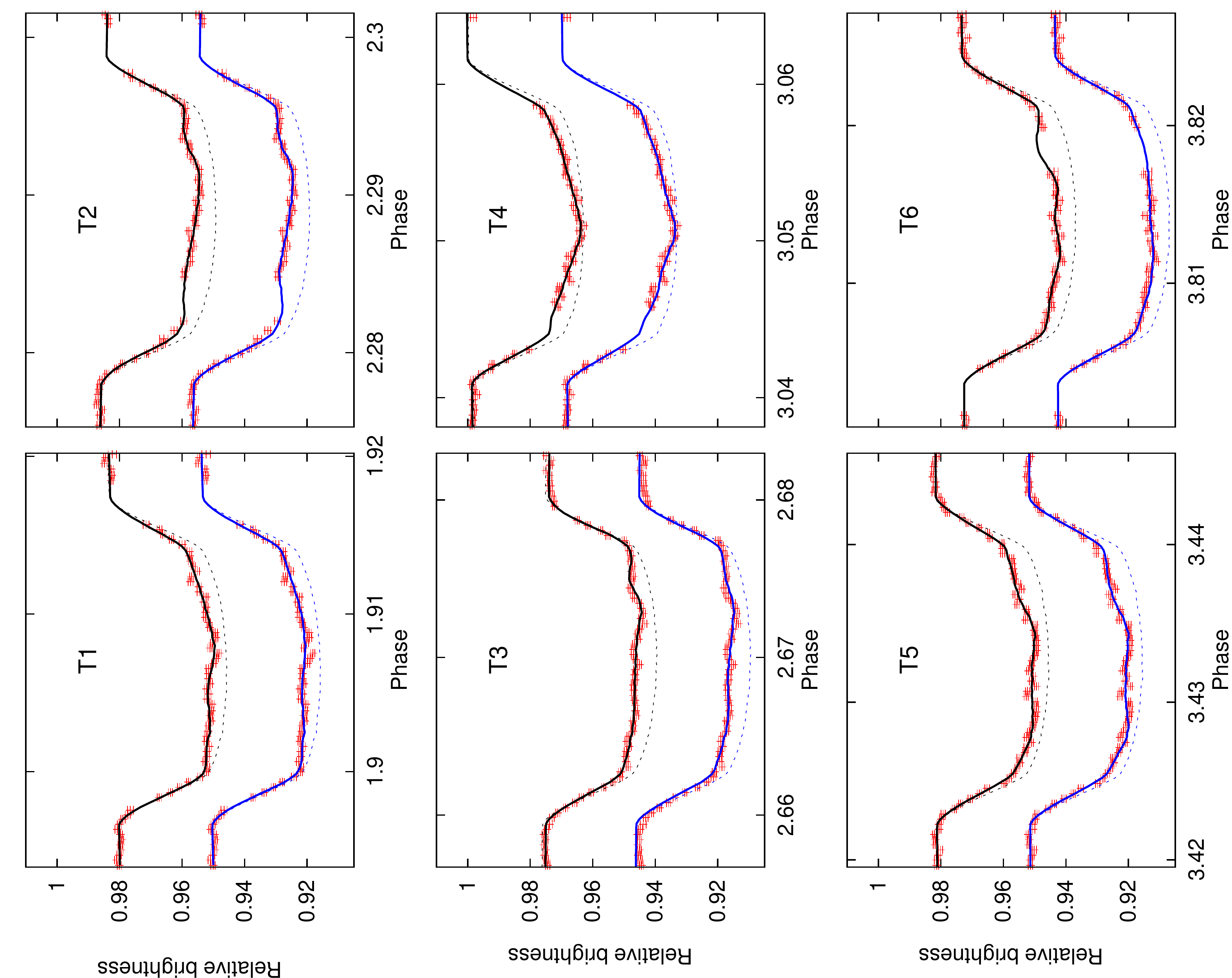}
    \caption{Close-up on the six individual transits ($128$~s binning).
             Observations are drawn as red points (including $1\sigma$~errors).
             The upper curve (black) in each panel shows the fit obtained from 1~rotation
             (phases $1.85$ to $2.85$ for T$1-3$ and $2.85$ to $3.85$ for T$4-6$);
             the lower curve (blue) gives the fit obtained by modeling both rotations (as seen in Fig.~\ref{fig:modelLC_global}).
             The dashed lines show the undisturbed transit profile for comparison.
    \label{fig:modelLC_transit}}
  \end{figure}

  In Figs.~\ref{fig:modelLC_global} and \ref{fig:modelLC_transit} we present the results of our modeling.
  Figure~\ref{fig:modelLC_global} shows the entire sub-sample of CoRoT data points
  used in our modeling as well as our lightcurve model in the upper panel.
  In the lower panel we show the model residuals (see Sect.~\ref{Sec:Analysis} for the definition of the
  error).
  Obviously, the data are matched well; however, there are systematic offsets between
  the observation and the model. In particular, the model tends to overestimate the observations during the first half of the time span,
  whereas it underpredicts it in the second half. This effect is related to surface evolution already detectable on time scales
  below the rotation period \citep{Lanza2009} (also see Sect.~\ref{Sec:RotPer}).
  Within the transits the residuals remain small compared to the rest of the lightcurve.
  This must be regarded a consequence of both the smaller bin size of $128$~s used here and
  also the twofold better resolution of the model on the eclipsed section. Note that during the fit the transit residuals are
  `overweighted' by a factor of ten to avoid them to be prevailed by the much larger global residuals.
  Although, the deviations can be as large as $10\,\sigma$, the mean deviation of the global lightcurve from the model
  amounts to $620\times 10^{-6}$ not far from the theoretical limit of $560\times 10^{-6}$ (cf. Sect.~\ref{Sec:RotPer}).

  The lightcurve presented in Fig.~\ref{fig:modelLC_global} contains six transits (labeled \mbox{`T$1-6$')}.
  The associated transit lightcurves together with our models are shown in detail in Fig.~\ref{fig:modelLC_transit}.
  Each individual panel shows the same transit twice: The lower curve represents a transit reconstruction from the
  full data sample (phases \mbox{$1.85-3.85$)}, and the upper curve denotes a reconstruction from only the first (T$1-3$)
  or second \mbox{(T$4-6$)} half of the sample data (shifted up by $0.03$).
  The dotted lines show the transits as we would observe them without any activity on the eclipsed section of the surface, where we
  assume a brightness of $1$ for the underlying photosphere.
  
  The transit reconstructions obtained from half of the sample data reproduce the transit substructure very accurately.
  The resulting surface reconstructions are, however, unreliable where the surface is insufficiently covered
  (cf. Fig.~\ref{fig:PhVis}, around longitudes of $180$\td \ and $320$\td).
  Interestingly, those reconstructions based on data from two rotation phases also recover most of the transit substructure and
  are by no means off the mark.
  When both rotation periods are used, $\chi^2$ typically increases by $10-20$~\%, a difference
  hardly visible in Fig.~\ref{fig:modelLC_transit}. As an exception, the fit quality of the third transit (T$3$), decreases dramatically, with
  $\chi^2$ increasing by a factor of $\approx 2.5$. This is, however, mainly a consequence of the observed surface evolution shifting the
  continuum level.
  The overall stability of the fit quality indicates that lifetimes of surface features are of the order of a few stellar rotation, which is in
  agreement with the results of \citet{Lanza2009}.
  
\begin{figure}[h]
    \includegraphics[angle=-90, width=0.49\textwidth]{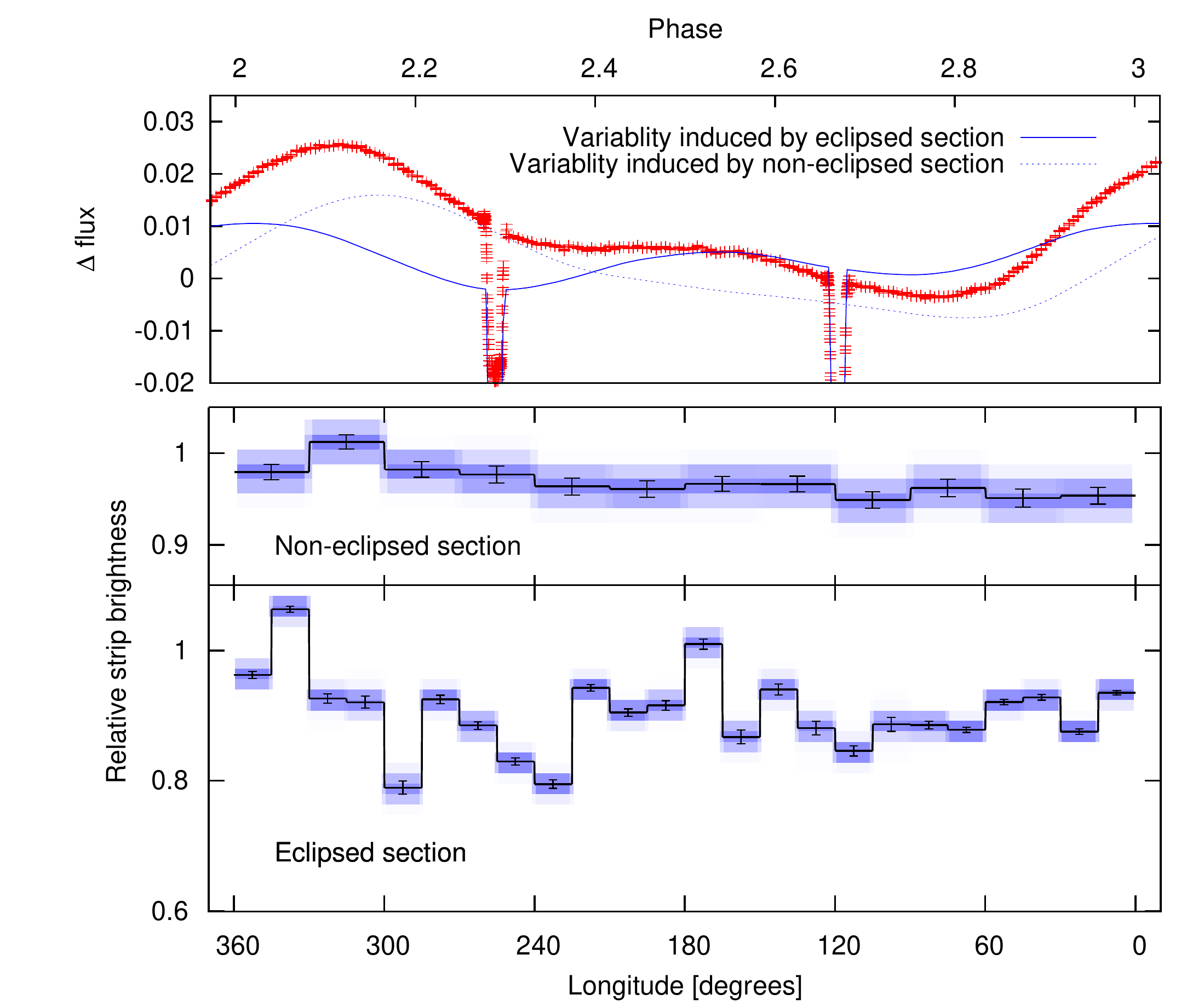}
    \caption{
    Upper panel: The CoRoT data (red symbols) and the flux modulation contributed in our model
    by the eclipsed (blue curve) and non-eclipsed section (dotted gray curve) individually.
    All curves are median subtracted.
    Middle and lower panel: Brightness distribution of the strips located on the eclipsed and non-eclipsed section of the surface.
    The color gradient renders the distribution obtained from
    $50$~reconstructions with randomized starting points.
    \label{fig:modelSurface}}
  \end{figure}

  In Fig.~\ref{fig:modelSurface} (lower and middle panel) we present the strip brightness distribution pertaining to the
  lightcurve model shown in Figs.~\ref{fig:modelLC_global} and \ref{fig:modelLC_transit},
  i.e., a 1D-reconstruction of the surface.
  We estimated mean and errors by recording the distribution of the parameter values obtained from
  $50$~reconstructions with randomized starting points, and the respective distributions are indicated by the color gradients in
  Fig.~\ref{fig:modelSurface}. The error bars correspond to the associated standard deviations. They
  reflect the ability of the fitting algorithm to converge to a unique extremum, which is determined by both the characteristics
  of the algorithm and the structure of the fit statistics.
  Investigating the brightnesses distribution of the non-eclipsed strips,
  we notice a slight degeneracy in some of the $12$~non-eclipsed strips,
  i.e., a fraction of the brightness may be redistributed without considerable loss of fit quality.
  The averaging of the $50$~reconstructions flattens out such features, and, thus, acts like a regularization
  of the brightness distribution. No such effect is observed for the eclipsed strips.
  
  We compared our results to the reconstructions given by \citet{Lanza2009} (their Fig.~4) and find
  our longitude scale to be shifted by $\approx 70$\td \ in respect to the \citeauthor{Lanza2009} scale.
  Our reconstructions show the same bright band at a longitude of $\approx 260$\td \ ($330$\td \ in our work).
  Tentatively averaging over an appropriate `time band' in their Fig.~4, we also find qualitative agreement for the
  remaining spot distribution.
  
  \begin{figure}[h]
     \includegraphics[width=0.45\textwidth]{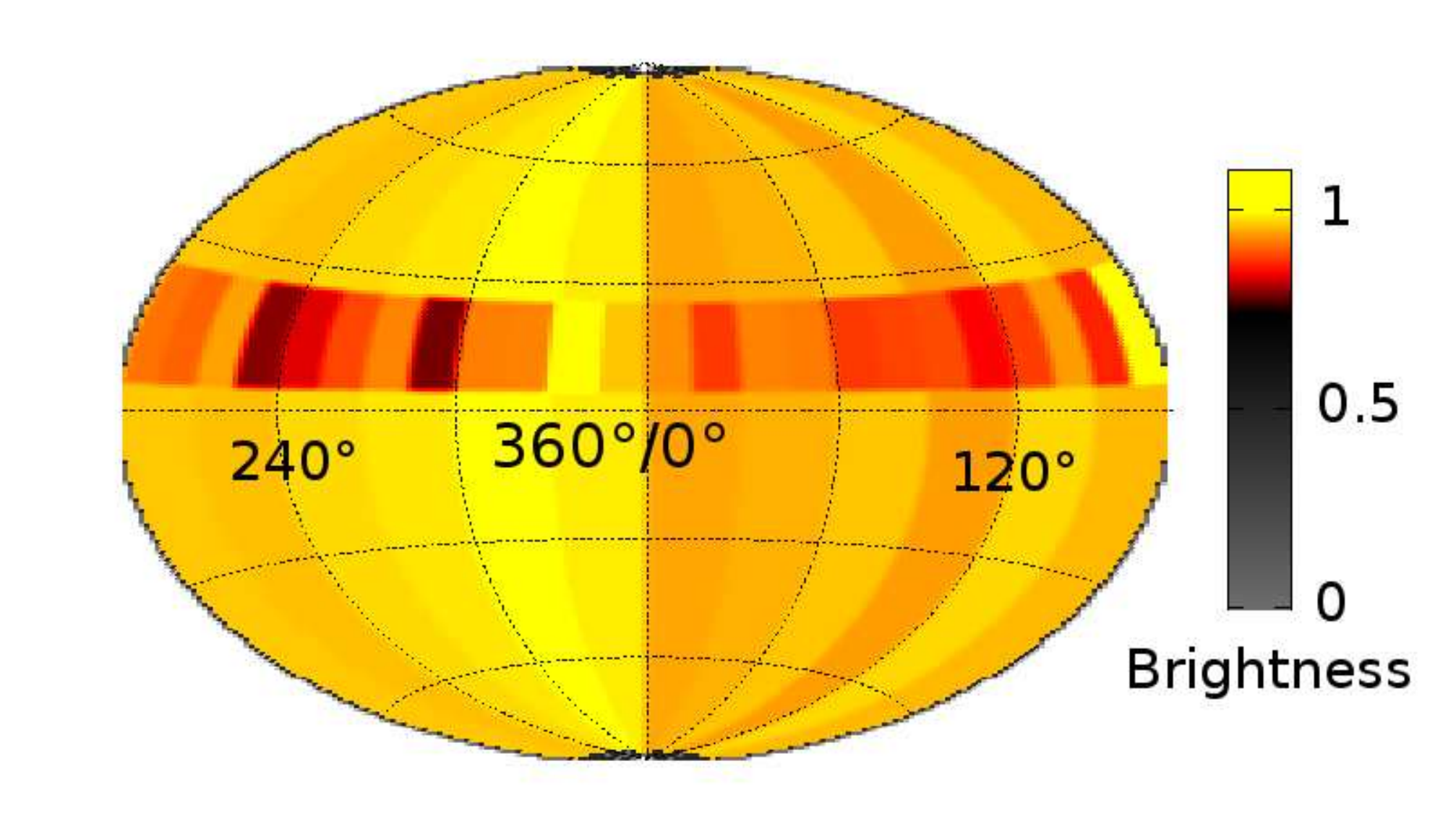}
     \caption{Surface map of CoRoT-2a showing the reconstructed brightness distribution.
              Spots located on the non-eclipsed surface are, due to their unknown latitude,
              blurred over the entire reconstruction strip
              resulting in the lower contrast compared to the eclipsed section.
     \label{fig:StripMap}}
  \end{figure}
  
  Clearly, the flux fraction contributed by the eclipsed strips is smaller than that of the
  non-eclipsed strips, because they are smaller by a factor of five.
  In the upper panel of Fig.~\ref{fig:modelSurface} we show
  the lightcurve model contributions provided by the eclipsed and the non-eclipsed section with their sum making up the
  model for the CoRoT data which is also shown.
  The median flux level was subtracted from all curves to emphasize the modulation amplitude in favor of flux level.
  Obviously, the modulation amplitudes induced by the eclipsed and non-eclipsed section approximately balance.
  This indicates that their influence on the stellar variability is of the same order despite their large
  difference in size.

  A visualization of our best fit surface reconstruction is presented in Fig.~\ref{fig:StripMap},
  showing a Hammer projection of the associated distribution of surface brightnesses.
  The planet-defined low latitude band shows especially dark features e.g. in the range of $200$\textdegree \ to
  $300$\textdegree \ in longitude and is clearly visible.
  Also the non-eclipsed sections of the star show significant variations.
  Note that our map only shows the average brightness of these regions;
  since the non-eclipsed regions are larger by area, they contribute more flux, however,
  the `missing' flux in these regions is likely also concentrated in spots.
  In the following section we address the issue of the flux contribution from the eclipsed and non-eclipsed sections.

 \subsubsection{Brightness distribution and spot coverage}
   
   Without a very precise absolute flux calibration \citep[as e.g. in][]{Jeffers2006},
   lightcurve analyses can usually only investigate the inhomogeneous part of the entire spot coverage. This statement is, however,
   partially invalidated by a transiting planet because it breaks the symmetry of the problem: Spots being eclipsed by the planetary
   disk distort the transit profiles no matter whether they belong to a structure which appears symmetric on a global scale or not.
   
   As an example, assume that half the eclipsed section of \co-2a, say longitudes $0-180$\textdegree, is spotted while the other half is
   covered by undisturbed photosphere.
   Clearly, the transits will be shallower when the planet eclipses the dark portion of the star, and they will be deeper
   when the bright section is eclipsed. Also the lightcurve will be distorted. Now further assume that a comparable section between
   longitudes $180-360$\textdegree \ is dark on the opposite hemisphere of the star outside the eclipsed band.
   In this case the global spot configuration is
   perfectly symmetric with respect to longitude and the global lightcurve does not show any trace of activity. Yet, the transits will
   still be shallower when the planet eclipses the dark band, and they will still be deeper when the bright surface is eclipsed.
   
\begin{figure}[h]
    \includegraphics[angle=-90, width=0.49\textwidth]{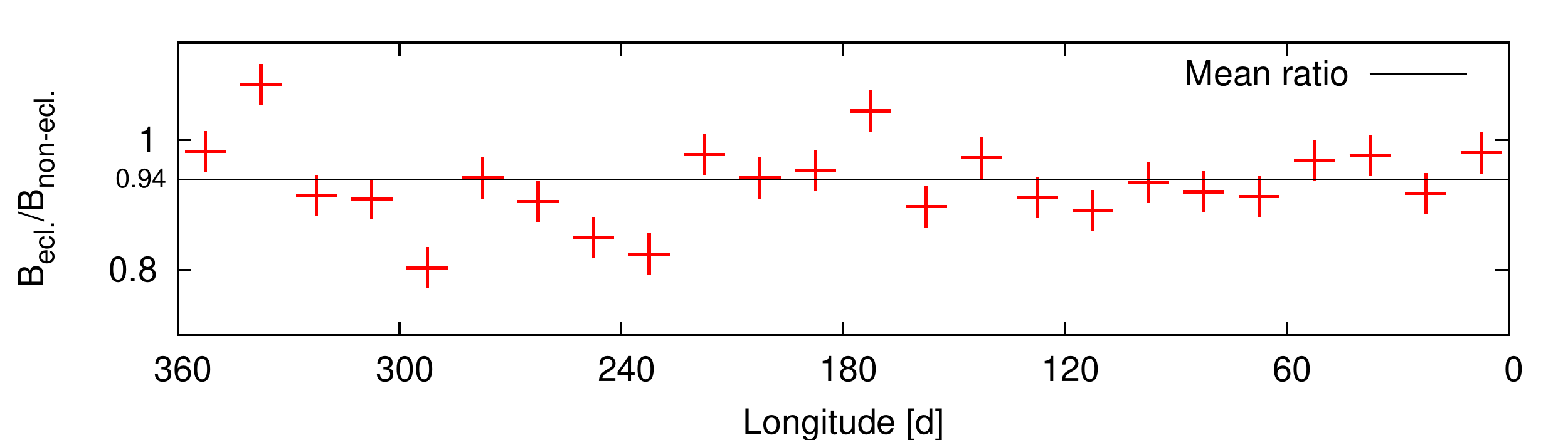}
    \caption{Ratio of the surface brightness in the eclipsed ($B_{ecl.}$ ) and non-eclipsed ($B_{non-ecl.}$) section.
    \label{fig:SpotDenRat}}
  \end{figure}

   If the spots were distributed symmetrically across the stellar surface, we would expect the stellar surface to be homogeneously
   bright. In Fig.~\ref{fig:SpotDenRat} we show the brightness ratio of eclipsed and non-eclipsed section as a function
   of longitude. Since there are more eclipsed than non-eclipsed strips, we always compare strips
   covering the same longitude.
   Only in two cases the eclipsed section is brighter than its non-eclipsed counterpart, while in $22$~cases it is not.
   
   The mean ratio is $0.94\pm0.01$ so that the part of the star passingly covered by the planet is found
   to be $6$~\% darker than the rest of the surface.
   Note that the remaining (non-transited) surface is brighter \textit{on average}, locally it may even be darker.

\section{Discussion and conclusions}
\label{Sec:Dis}
We present a surface reconstruction for the planet host star CoRoT-2a.
Our modeling is based on a CoRoT data interval covering two full stellar rotations,
and it treats the entire lightcurve -- including the planetary transits -- in a consistent way.

We show that a consistent modeling of the lightcurve is possible using a static model,
i.e., not including any spot evolution.
Although surface evolution on scales of the stellar rotation period is seen
in both the reconstruction of the global lightcurve \citep[as already reported by][]{Lanza2009} and the transit lightcurves,
this effect is small in the context of our analysis.
The static model provides reasonable fits to six consecutive transit lightcurves.
The associated surface configuration changes little during this period, and,
therefore, the surface evolution must be relatively slow compared to the time scale of $\approx 9$~d under consideration.
This time scale is also valid for the lifetimes of spots on the eclipsed surface section.

Our results indicate that the planet-eclipsed band on the stellar surface is -- on average --
about $6$~\% darker then the remaining part of the surface.
\citet{Lanza2009} note that the strength of differential rotation derived from their lightcurve fits
seems very low in comparison to expected values derived from measurements in other systems \citep{Barnes2005}.
They speculate that this may indicate a spot distribution limited to a narrow latitude band.
If this should be true,
the latitude band is possibly located at low latitudes,
i.e., within $\pm 30$\td \ around the equator as observed on the Sun.
In this case it covers the eclipsed section where we find a darker surface, i.e., higher spot coverage.
We caution that this result may also be influenced by the adopted planetary parameters (mainly the size),
which are hard to determine accurately \citep{Czesla2009}.

We checked whether the effect of gravitational darkening could significantly contribute to a darker surface in the vicinity of the equator.
For the stellar parameters of \co-2a, we find that the (effective) gravitational
accelerations at the poles and at the equator are equal to within $0.07$~\%, so that gravitational darkening does not provide
a significant contribution to the brightness gradient found in our modeling;
this result is nearly independent of the assumed
coefficient, $\beta_1$
(\mbox{$T_{eff}^4\,\wasypropto\,g^{\beta_1}$} with the effective surface temperature $T_{eff}$
and the surface gravity $g$),
which is of the order of $0.3-0.4$ for \co-2a \citep{Claret2000}.

The `narrow-band hypothesis', i.e., a higher spot coverage in the planet-eclipsed
section compared to the non-eclipsed surface,
also provides a natural explanation for the fact that both the eclipsed and non-eclipsed
surface regions account for about the same amplitude of variation in the lightcurve.
Using the Sun as an analogy again, we would qualitatively expect the same structure, as seen under the planet path,
on the opposite hemisphere as well;
two `active belts' that are symmetric with respect to the equator.
The non-eclipsed activity belt, which would be only observable in the global lightcurve,
would then be primarily responsible for the variability of the lightcurve
contributed by the non-eclipsed surface section.

We conclude that our results support a surface model consisting of active regions north and south of the equator,
possibly even bands of spots at low latitudes analogous to the Sun.
Further investigations of this system using more sophisticated models (first of all surface evolution)
and using the entire observation interval of approximately $140$~days have the potential to reveal more information
on the constantly changing surface distribution of spots on CoRoT-2a.

%

\begin{acknowledgements}
K.H. is a member of the DFG Graduiertenkolleg 1351 \textit{Extrasolar Planets and their Host Stars}.
S.C. and U.W. acknowledge DLR support (50OR0105).
\end{acknowledgements}


\bibliographystyle{aa}
\bibliography{referenz}


\end{document}